\documentclass[aps,pra,superscriptaddress,twocolumn,showpacs]{revtex4-1}

\pdfoutput=1

\usepackage{color}
\usepackage{amsmath, amsthm, amsfonts}    
\usepackage{amssymb}
\usepackage{mathptmx} 
\usepackage{dsfont}
\usepackage{tikz}
\usepackage{amsmath}

\usepackage[unicode=true,pdfusetitle, bookmarks=true,bookmarksnumbered=false,bookmarksopen=false, breaklinks=true,pdfborder={0 0 0},backref=false,colorlinks=true,citecolor=blue]{hyperref}

\usepackage{graphicx}   
\usepackage[caption=false]{subfig}       
\usepackage{bm}            
\usepackage[normalem]{ulem}  

\newlength\imagewidth
\newlength\imagescale

\def\be{\begin{eqnarray}}
\def\ee{\end{eqnarray}}

\def\im{{\rm i}}

\begin{document}
\title{Nonlocal magnetolectric effects in diffusive conductors with  spatially inhomogeneous  spin-orbit coupling}

\author{Cristina Sanz-Fern\'andez}
\email{cristina\_sanz001@ehu.eus}
\affiliation{Centro de F\'{i}sica de Materiales (CFM-MPC), Centro Mixto CSIC-UPV/EHU, 20018 Donostia-San Sebasti\'{a}n, Spain} 

\author{Juan Borge}
\affiliation{Nano-Bio Spectroscopy Group, Departamento de F\'isica de Materiales, Universidad del Pa\'is Vasco (UPV/EHU), 20018 Donostia-San Sebasti\'{a}n, Spain} 

\author{Ilya V. Tokatly}
\affiliation{Nano-Bio Spectroscopy Group, Departamento de F\'isica de Materiales, Universidad del Pa\'is Vasco (UPV/EHU), 20018 Donostia-San Sebasti\'{a}n, Spain} 
\affiliation{IKERBASQUE, Basque Foundation for Science, 48011 Bilbao, Spain}
\affiliation{Donostia International Physics Center (DIPC), 20018 Donostia-San Sebasti\'{a}n, Spain}

\author{F. Sebasti\'an Bergeret}
\email{sebastian\_bergeret@ehu.eus}
\affiliation{Centro de F\'{i}sica de Materiales (CFM-MPC), Centro Mixto CSIC-UPV/EHU, 20018 Donostia-San Sebasti\'{a}n, Spain}
\affiliation{Donostia International Physics Center (DIPC), 20018 Donostia-San Sebasti\'{a}n, Spain}

\begin{abstract}
We present a theoretical study of non-local magnetoelectric effects in diffusive hybrid structures with an intrinsic linear-in-momentum spin-orbit coupling (SOC) which  is assumed  to be spatially inhomogeneous. Our analysis is based on the SU(2)-covariant drift-diffusion equations from which we derive boundary conditions at hybrid interfaces for  SOC of any kind.  
Within this formulation, the spin current is covariantly conserved when the spin relaxation is only due to the intrinsic SOC. This conservation leads to the absence of spin Hall (SH) currents in homogeneous systems. If, however, extrinsic sources of spin relaxation (ESR), such as magnetic impurities and/or a random SOC at non-magnetic impurities, are present the  spin is no longer covariantly conserved,  and SH currents appear. 
We apply our model to describe non-local transport in a two-dimensional system with an interface separating  two regions: one normal region without intrinsic SOC and one with a Rashba SOC. We first explore the inverse spin-galvanic effect, \textit{i.e.}, a spin polarization  induced by an electric field. We demonstrate how the spatial behavior of such   spin density depends on both, the direction of the electric field and the strength of the ESR rate.  We also study the spin-to-charge conversion, and compute the  charge current and the distribution of electrochemical potential in the whole system when  a spin current is injected  into the normal region.   In systems with an inhomogeneous SOC varying in one spatial  direction,  we find an interesting non-local reciprocity between  the  spin density induced by a charge current at a given point in space,  and the spatially integrated current induced by a spin density injected at the same point.
\end{abstract}

\maketitle

\section{Introduction}

Spin-orbit coupling (SOC) in metals and semiconductors couples the charge and spin degrees of freedom of the electrons and leads to a variety of  magnetoelectric effects.   For that reason, conductors with sizable SOC are used for the creation and control of spin currents and spin densities  by applying electric fields.  Reciprocally, magnetoelectric effects allow for detecting spin by measuring electric signals~\cite{Zutic04,sinova2015spin}.

It is customary to distinguish between two kinds of magnetoelectric effects mediated by SOC: those relating spin and charge currents (spin Hall effect and its inverse), and those relating spin polarization and charge current (spin-galvanic effect and its inverse). The spin Hall effect (SHE) is the generation of a spin current, transverse  to the applied charge current~\cite{hirsch1999spin, dyakonov2017spin, Vignale2010, sinova2015spin}. The inverse effect, commonly known as the inverse SHE~
\cite{sinova2015spin}, corresponds to the spin-to-charge counterpart and consists of a charge current,  or a Hall  voltage, induced by a given spin current. Both direct and inverse SHE have been measured in several experiments and different materials~\cite{Kato04, Sih05,valenzuela2006direct,Takahashi_Revese_PRL07,Takahashi_GSH_NatMater08,Sagasta16}.

Here, we focus  on the charge-to-spin conversion due to the spin-galvanic effect (SGE), which refers to the generation of a charge current by creating a non-equilibrium spin polarization in the material. It takes place, for example, in materials with a linear-in-momentum intrinsic SOC, such as the Rashba or linear Dresselhaus SOC~\cite{vasko1979spin,bychkov1984oscillatory, dresselhaus1955spin}. Conversely, the inverse SGE corresponds to the spin polarization induced by applying an electric field/current~\cite{dyakonov1971current, Lyanda-Geller89, aronov1991spin,zhang2015spin,bel2008magneto,golub2011spin,luengo2017current, gorini2017theory}, which in the particular  case of Rashba SOC, is also known as the Edelstein effect~\cite{Edelstein1990, Shen2014}. In contrast to the SHE, the induced spin is homogeneous in space and, in principle, in the stationary case, no spin currents are generated~\cite{mishchenko2004spin, raimondi2005spin, inoue2004suppression, raimondi2006quasiclassical, ol2005spin}. Observation of the SGE and its inverse  has been reported in Refs.~\cite{Sih05,Stern06,ganichev2002spin,sanchez2013spin,Karube2016,isasa2016origin}.

\begin{figure}
    \centering
    \includegraphics[width=1\columnwidth]{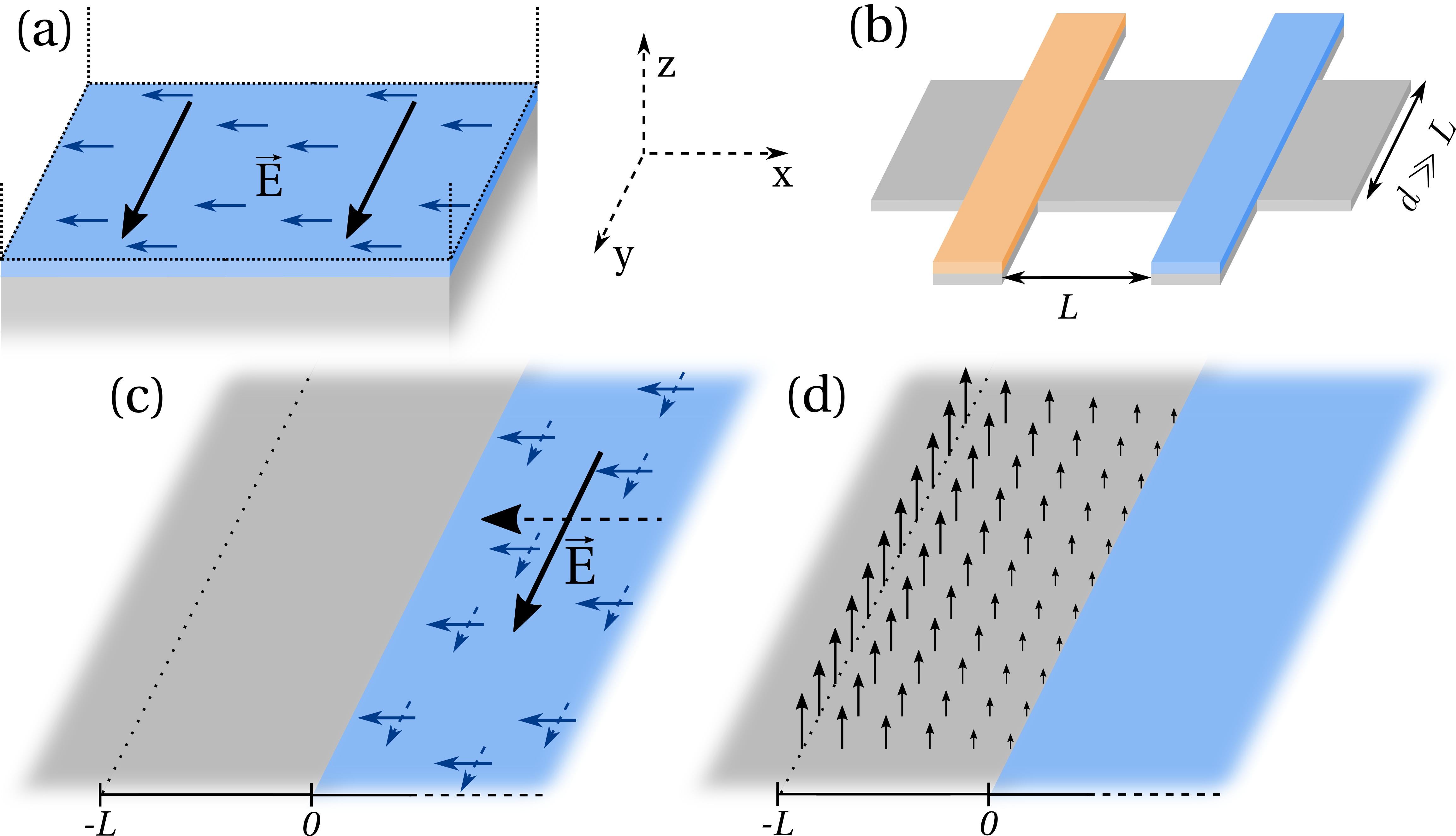}
    \captionsetup{justification= raggedright}
    \caption{Schematical view of different setups considered in this work. (a) Heterostructure with a localized SOC at the interface (blue region) between two different materials, in this case a normal conductor (gray bottom region) and an insulator (transparent top region). (b) Hybrid lateral structure for non-local transport measurements.  The gray region is  a normal conductor without SOC  connected to two electrodes. At the contact with the  blue electrode  it is assumed a sizable SOC, whereas the orange electrode is a  ferromagnet which may serve as a spin injector or detector.  (c) The system under consideration to study the non-local inverse SGE.  An electric  field is applied parallel or perpendicular  to the interface, between a normal conductor without SOC (gray region) and a conductor with intrinsic SOC (blue region). A spin density is induced in the latter. We are interested in the  spin density at a distance $L$ away from the interface in  the normal conductor. (d) Same setup as panel (c) but now a spin is injected into the normal conductor at a distance $L$ from the interface. We are interested in the charge current induced in the blue region as a consequence of the SGE.}
    \label{fig_sketch}
\end{figure}

From the experimental point of view, hybrid structures combining different materials play an important role in the detection of magnetoelectric effects.  This  requires,  on the one hand,   materials with large SOC for an efficient charge-to-spin conversion and, on the other hand, large enough spin diffusion lengths in order to transport the spin information across the device. 
At first glance, it seems difficult to find systems satisfying these two conditions, because a strong SOC in a diffusive system will inevitably lead to a strong spin relaxation~\cite{dyakonov1971current}. This problem can be overcome by using hybrid structures combining, for example, two different materials, one with a strong SOC, in which the charge-to-spin conversion occurs, adjacent to a second material  with a weak SOC where the spin information can be transported over  long distances. This conversion can also take place at the interface between a metal and an insulator with a sizable SOC [Fig.~\ref{fig_sketch}(a)]~\cite{kim2017evaluation}.

An efficient way of injecting and detecting the spin accumulation is by using  non-local spin valves, as the one sketeched in  Fig.~\ref{fig_sketch}(b)~\cite{jedema2001electrical,jedema2002electrical,johnson1985interfacial, isasa2016origin,safeer2019room, ghiasi2019charge, li2019electrical, benitez2019tunable, hoque2019all}. In this setup the source signal, either a spin or a charge current, is injected from one of the electrodes (orange/blue), whereas the response signal, a charge or spin voltage, is measured non-locally at the detector electrode (blue/orange). Similar valves combining ferromagnetic electrodes and metallic wires have also been used to measure the SHE~\cite{valenzuela2006direct}, as well as to study the  reciprocity between the SHE and the inverse SHE~\cite{niimi2015reciprocal}.

In this work, we present a theoretical study of non-local electronic transport in hybrid diffusive systems with linear-in-momentum intrinsic SOC of any type. We focus on the reciprocity between the non-local SGE and its inverse. Our analysis is based on the drift-diffusion equations~\cite{dyakonov2017spin} formulated in the language of SU(2) gauge fields, where  the  intrinsic SOC and the  Zeeman field  enter  as the space and  time  components of an effective SU(2) 4-potential~\cite{gorini2010non,tokatly2008equilibrium,tokatly2017usadel}. Within this formalism, the spin obeys a covariant continuity equation which  explains  the absence of spin Hall (SH) currents in a homogeneous system with intrinsic SOC. This covariant conservation of the spin  is broken in the presence of any extrinsic source of spin relaxation (ESR), as for example magnetic impurities or a random SOC originated from scattering of electrons at non-magnetic impurities. Such symmetry breaking leads to a finite spin current that may flow into a material without SOC.

In order to describe the transport in such hybrid systems, we derive effective boundary conditions (BC) valid for systems of any dimension. These BC describe the transport between diffusive conductors with different (not only in the strength) linear-in-momentum SOC and mean-free path, and they are valid for any direction of the applied field~\footnote{Our analysis is done within the diffusive limit and hence it is assumed that the spatial variation of the SOC occurs over a length scale larger than the momentum relaxation length. In this regard, we do not take into account so-called, interfacial SOC, i.e., SOC that only exists over atomic lengths right at the interface (see, for example, Refs.~\cite{borge2017ballistic, borge2019boundary, amin2018interface}).}.  They generalize the BC obtained in Refs.~\cite{tserkovnyak2007boundary,raimondi2006quasiclassical,adagideli2007extracting,brataas2007spin, wang2011rashba}, for a specific case of 2D Rashba systems.

We apply the diffusion equation and BC  to study  non-local measurement  of the SGE and its inverse in a two-dimensional hybrid system consisting of a diffusive conductor without intrinsic SOC, labeled as normal conductor, adjacent to a Rashba conductor, i.e., a conductor with an intrinsic Rashba SOC [see Figs.~\ref{fig_sketch}(c) and (d)]~\cite{safeer2019room, ghiasi2019charge, li2019electrical, benitez2019tunable, hoque2019all}. 
Firstly, we address the non-local inverse SGE [Fig.~\ref{fig_sketch}(c)], and calculate  the value of the spin density induced 
at the normal conductor at a finite distance from the interface with the  Rashba region, when an electric field is applied.
If the field is parallel to the interface, and due to its covariant conservation, the spin generated at the Rashba conductor cannot diffuse into the normal region, leading to a zero signal~\cite{tserkovnyak2007boundary,raimondi2006quasiclassical,adagideli2007extracting,brataas2007spin}. However, inclusion of  ESR breaks the covariant conservation of the spin, and a finite SH current is generated. This  leads to a diffusion of the spin into the normal region. We emphasize that this, previously unnoticed, mechanism of the spin injection is different from the one appearing at the boundary between materials with different intrinsic SOC and different elastic mean-free paths~\cite{brataas2007spin}.
If  the electric field is applied  perpendicular to the interface, the situation is rather different. In this case, the BC impose the conservation of both the spin density and spin (diffusive) currents at the interface. This leads to a diffusion of the spin density induced via the inverse SGE into the normal conductor even in the absence of ESR mechanisms. For the specific case of a Rashba SOC, this result coincides with the one of Ref.~\cite{brataas2007spin}.

Secondly,  we address the inverse effect, i.e., the non-local SGE [see Fig.~\ref{fig_sketch}(d)]. In this case, a spin density  is injected into the normal conductor at a certain distance from the interface. This spin diffuses over the normal region, and the corresponding spin diffusion current reaches the Rashba conductor, where it is transformed into a charge current.
We demonstrate that depending on the polarization of the injected spin density, charge currents parallel or perpendicular to the interface can be generated.  
In the absence of  ESR, the spatially integrated  current parallel to the interface vanishes, leading to zero global SGE, whereas a finite ESR   leads to a finite total charge current. 
These two situations are  the reciprocal  to the  non-local  inverse SGE described above.  Indeed, we find  a hitherto unknown general  non-local reciprocity relation between the global charge current induced by the spin injected locally at some point $x_0$, 
and the spin at the same point $x_0$ induced by applying a uniform electric field. 

Finally, we consider the non-local SGE when the lateral dimensions are finite.  In this case, no current  flows at the lateral boundaries.  We solve this boundary problem and find a redistribution of the electrochemical potential and the local charge currents  flowing  in the whole structure.

The paper is organized as follows. In Sec.~\ref{sec_model} we review the SU(2)-covariant drift-diffusion equations for  the  charge and the  spin densities, and  derive the  general BC. In Sec.~\ref{sec_cs} we  first review  the  inverse SGE   in bulk homogeneous systems with linear-in-momentum SOC.  In Sec.~\ref{sec_inv_SGE} we explore the non-local inverse SGE in the normal/Rashba conductor structure shown in Fig.~\ref{fig_sketch}(c). In Sec.~\ref{sec_rec} we analyze at a general level  the reciprocity between this effect and the non-local SGE, and show that the spin density at $x_0$, induced via the non-local inverse SGE, is proportional to the integrated charge current, generated via the non-local SGE, by injection of a spin current density at $x_0$.  
In Sec.~\ref{sec_sc} we study the non-local SGE shown in Fig.~\ref{fig_sketch}(d) and compute the spatial dependence of the charge current density. In Sec.~\ref{sec_red} we analyze the SGE in the same structure but with finite lateral dimensions, and determine the redistribution of charge currents and the electrochemical potential induced by the SGE. Finally, we present our conclusions in Sec.~\ref{sec_conclusions}. 

\section{Theoretical description of  diffusive hybrid structures}
\label{sec_model}
In this section we review  the diffusion equations for the charge and spin densities in homogeneous systems in the SU(2)-covariant formalism. We derive BC for these equations at hybrid interfaces.  These conditions describe interfaces between materials with different kinds and strengths of linear-in-momentum SOC.

We focus on  materials with linear-in-momentum SOC. The latter  can be described with the help of a SU(2)  vector potential $\hat{\mathcal{A}}_k = \frac{1}{2} \mathcal{A}_k^a \sigma^a$. Specifically, we consider the following Hamiltonian~\cite{raimondi2012spin,tokatly2017usadel}:
\begin{equation}
H = \frac{(p_k - \hat{\mathcal{A}}_k)^2}{2m} - \hat{\mathcal{A}}_0 + V_{\rm{imp}}\; . \label{eq:Hamiltonian1}
\end{equation}
A particular case of SOC is the  widely studied two-dimensional Rashba SOC, for which  $\mathcal{A}_y^x = - \mathcal{A}_x^y = 2m\alpha = \lambda_{\alpha}^{-1}$, being $\alpha$ the Rashba parameter.  The second term in Eq.~\eqref{eq:Hamiltonian1},  $\hat{\mathcal{A}}_0 = \frac{1}{2} \mathcal{A}_0^a \sigma^a$, describes a Zeeman or ferromagnetic exchange field. The last term is the potential of randomly distributed impurities. We consider both non-magnetic and magnetic impurities, such that $V_{\rm{imp}} = V_{\rm{nm}} + V_{\rm{m}}$, where $V_{\rm{nm}}$ contains the SOC generated by the random potential of the impurities~\cite{Zutic04}.  In our notation, lower indices correspond to spatial coordinates and upper indices to spin components, and  throughout the paper the summation over repeated indices is implied. 

The advantage of introducing the SOC as a SU(2) gauge  field, is that one can derive a  SU(2)-covariant continuity equation for the spin~\cite{tokatly2008equilibrium}. 
In other words,  within this formalism the spin is covariantly conserved when only intrinsic linear-in-momentum SOC is considered, and it satisfies  the following continuity equation:
\begin{equation}
\label{eq_cont}
\tilde\partial_t \hat S + \tilde{\partial}_k {\hat{j}}_k =0\; ,
\end{equation}
where $\hat S = S^a(\sigma^a/2)$ is the spin density and $ \hat{j}_k = j_k^a(\sigma^a/2)$ is the spin current density flowing in the $k$ direction, defined as the average of the spin current operator, $j^a_k=(1/2)\left\{\partial H/\partial p_k, \sigma^a/2 \right\}$. The spin continuity equation has the same form as  the charge continuity equation,  but with the derivatives substituted by the covariant ones, $\tilde{\partial_t} = \partial_t - i [\hat{\mathcal{A}}_0, \cdot]$ and  $\tilde{\partial_k} = \partial_k - i [\hat{\mathcal{A}}_k, \cdot]$, respectively. 

In the presence of any kind of ESR, as for example magnetic impurities, or SOC due to the impurity scattering~\cite{raimondi2012spin,Shen2014,huang2018extrinsic}, Eq.~\eqref{eq_cont} acquires an additional term:
\begin{equation}
\tilde\partial_t \hat S + \tilde{\partial_k} \hat{j}_k =- \frac{1}{\tau_{\rm{ext}}} \hat{S}\; . \label{eq_cont2}
\end{equation}
Here, we assume  that the spin relaxation is isotropic in space and neglect the interference term between extrinsic and intrinsic SOC~\cite{gorini2017theory}.
Clearly, the ESR breaks the SU(2) symmetry and hence the spin is no longer covariantly conserved. 

We now consider a hybrid interface between two materials with different mean-free path and SOC. 
In real systems, all potentials appearing in the Hamiltonian of Eq.~\eqref{eq:Hamiltonian1} must  be  finite and, therefore, both the spin currents and  spin densities must also be finite at any point in space. 
Therefore, one can integrate Eq.~\eqref{eq_cont2} over an infinitesimal interval across the interface and obtain the conservation of the spin current:
\begin{equation}
\label{eq_bc_j}
n_k j_{k}^a|_{0^+}\ = n_k j_{k}^a |_{0^-}\; ,  
\end{equation}
where ${\bf{\hat{n}}}=(n_x,n_y,n_z)$ is a unit vector perpendicular to the interface. This is the first BC. 

In order to describe the spatial distribution of the spin and charge densities, we focus here on diffusive systems in which 
the elastic  scattering rate  at non-magnetic impurities dominates over all other rates. Specifically, the inverse of the momentum relaxation time, $\tau^{-1}$, is assumed to be larger than all other energies, such as SOC, Zeeman field, or the inverse of any ESR time, $\tau_{\rm{ext}}^{-1}$. In this limiting case, the spin current is given by~\cite{gorini2015spin,tokatly2017usadel,konschelle2015theory}:
\begin{equation}
\hat{j}_k = -D \tilde\partial_k \hat S - \gamma \hat{\cal{F}}_{ki} j_i \; .
 \label{eq:gen_spin_current}
\end{equation}
The first term corresponds to the SU(2)-covariant diffusion current, where  $D = v_{\rm{F}}^2\tau/d$ is the  diffusion coefficient, and $d$ the dimension of the system.
The second term, proportional to the charge current density $j_i$, describes the charge-to-spin conversion, where $\gamma=\tau/(2m)$. It is, therefore, the term responsible for the SHE.
The proportionality factor contains  the field strength tensor defined in terms of the SU(2) vector potential as:
 \begin{equation}
\label{eq_Fij}
\mathcal{F}^a_{ij} = \partial_i \mathcal{A}_j^a - \partial_j \mathcal{A}_i^a + \mathcal{A}^b_i \mathcal{A}^c_j \varepsilon^{abc}\; .
\end{equation}
In analogy to the ordinary Hall effect, where electrons are deflected by an external magnetic field, the second  term in the right-hand side of Eq.~\eqref{eq:gen_spin_current} describes the spin-dependent deflection in the presence of an effective  SU(2) magnetic field, Eq.~\eqref{eq_Fij}, generated by SOC.

The charge current density in the diffusive limit is given by~\cite{gorini2015spin,tokatly2017usadel}:
\begin{equation}\label{eq:charge_current}
\begin{split}
{j}_k &= -D\partial_k n - \sigma_{\rm{D}} E_k - \gamma\hat{\mathcal{F}}_{ki}\hat{j}_i \; ,
\end{split}
\end{equation}
where $n$ is the out-of-equilibrium electron density, $\sigma_{\rm{D}}$ the Drude conductivity, and $E_k$ the $k$th component of the electric field. The third term is the 
reciprocal to the second term in Eq.~\eqref{eq:gen_spin_current}. It describes the spin-to-charge conversion under the action of the SU(2) field and, therefore, is related to the inverse SHE.

The spin and charge diffusion equations are obtained by substituting expressions Eqs.~\eqref{eq:gen_spin_current} and~\eqref{eq:charge_current} into Eq.~\eqref{eq_cont2} and the charge continuity equation, respectively. Specifically, 
the SU(2)-covariant spin diffusion equation in a stationary case and in the absence of a Zeeman field has the following compact form:
\begin{equation}
\label{eq_diff_charge0}
 \tilde\partial_k D\tilde\partial_k \hat{S} + \gamma \tilde{\partial}_k \hat{\mathcal{F}}_{ki} j_i = \frac{1}{\tau_{\rm{ext}}} \hat{S}\; .
\end{equation}
The covariant Laplace operator in the first term can be written explicitly by expanding the covariant derivatives~\cite{bergeret2014spin}:
\begin{equation}
\label{eq:exp_sp_derivative}
(\tilde\partial_k D\tilde\partial_k )^{ab} = \partial_k D\partial_k \delta^{ab} + 2 P^{ab}_k \partial_k + \partial_k P^{ab}_k - \Gamma^{ab}\; ,
\end{equation}
where we  define  the following operators~\cite{bergeret2014spin}:
\begin{eqnarray}
  &&P^{ab}_k = -iD[\hat{\mathcal{A}}_k, \cdot] = D\mathcal{A}^c_k \varepsilon^{cba}  \label{eq_Pab}\; ,\\
   &&\Gamma^{ab} = D[\hat{\mathcal{A}}_k, [\hat{\mathcal{A}}_k, \cdot]] =- D^{-1}P^{ac}_k P^{cb}_k  \label{eq_gamma}\; .
\end{eqnarray}
Here, $\Gamma^{ab}$ is the general Dyakonov-Perel relaxation tensor  that describes  spin relaxation 
due to the randomization of the spin precession  caused by  the scattering at non-magnetic impurities,
whereas $P^{ab}_k $  describes  the  precession of an inhomogeneous spin density~\cite{mishchenko2004spin, bergeret2014spin}. 

The spin diffusion equation~\eqref{eq_diff_charge0}, is solved in the next sections for different geometries and situations.  To describe hybrid interfaces between different materials, one  needs an additional BC to Eq.~\eqref{eq_bc_j}, which can be obtained by integrating Eq.~\eqref{eq:gen_spin_current} over a small interval around the interface. In the absence of charge current sources, this integration leads to the continuity of the spin density across the interface.
If a finite charge current density is  induced by an electric field, $j_i = -\sigma_{\rm{D}} E_i$,  we divide  Eq.~\eqref{eq:gen_spin_current}  by $D$ and integrate it over a small interval across the junctions and obtain:
\begin{equation}
\label{eq_bc_s}
\begin{array}{c}
     S^a \big|^{0^+}_{0^-}\! =\! \frac{1}{2}\! \left( \frac{\gamma\sigma_{\rm{D}}}{D} \big|_{0^+}\!+ \frac{\gamma\sigma_{\rm{D}}}{D} \big|_{0^-} \right)\! (\delta_{ij}-n_i n_j)\! \left( \mathcal{A}_{j}^a |_{0^+}\! - \mathcal{A}_{j}^a |_{0^-}\right)\! E_i . 
\end{array}
\end{equation}
In  this equation we allow for  different values of the momentum scattering time $\tau$ and SOC at  both sides of the junction, and different directions of the electric field with respect to the interface. Equation~\eqref{eq_bc_s} generalizes the result for 2D Rashba systems~\cite{tserkovnyak2007boundary,raimondi2006quasiclassical,adagideli2007extracting,brataas2007spin, wang2011rashba}, for any kind of linear-in-momentum SOC and any dimension [see Figs.~\ref{fig_sketch}(a) and (b)].

In the next sections we  study  non-local transport in the diffusive hybrid structure sketched in Figs.~\ref{fig_sketch}(c) and (d).
It  consists of a normal conductor without intrinsic SOC (gray area) adjacent to a conductor with Rashba SOC (blue area), from here on referred to as a {\it Rashba conductor}. As mentioned above, the Rashba SOC is described by the SU(2) vector potential with non-zero components $\mathcal{A}_y^x = - \mathcal{A}_x^y = 2m\alpha = \lambda_{\alpha}^{-1}$. In what follows, we assume that the momentum scattering time, $\tau$, is constant in the whole system, and focus on the effect of ESR. Furthermore, we assume that the system is invariant in the $y$ direction, such that the spin density only depends on $x$.

In the normal conductor region the spin current has only a diffusion contribution [first term of Eq.~\eqref{eq:gen_spin_current}], and  the spin diffusion equation, Eq.~\eqref{eq_diff_charge0}, has the same form for all spin components:
\begin{equation}\label{eq_diff_N}
    \partial_{x}^2 S^a = \frac{S^a}{\lambda_{\rm{s}}^2} \; ,
\end{equation} 
where $\lambda_{\rm{s}}$ is the spin diffusion length in the normal conductor. 

In the Rahsba conductor, the three components of the spin current $\hat{j}_x$ are obtained from Eq.~\eqref{eq:gen_spin_current}:
\begin{eqnarray}
   &&j_x^x = - D \partial_x S^x + \lambda_{\alpha}^{-1} S^z \label{eq_jx_E1}\; ,\\
   &&j_x^z = - D \partial_x S^z - \lambda_{\alpha}^{-1} S^x + \gamma \sigma_{\rm{D}}  \lambda_{\alpha}^{-2} E_y \label{eq_jx_E3}\; , \\
   &&j_x^y = - D \partial_x S^y \label{eq_jx_E2}\; .
\end{eqnarray}
The components of the spin density are determined by the following set of coupled diffusion equations:
\begin{eqnarray}
   &&\partial^2_x S^x = 2 \lambda_{\alpha}^{-1} \partial_x S^z + (\lambda_{\alpha}^{-2} + \lambda_{\rm{ext}}^{-2}) S^x - \frac{\gamma\sigma_{\rm{D}}}{D} \lambda_{\alpha}^{-3} E_y \label{eq_sde_1x}\; ,\\
   &&\partial^2_x S^z = - 2 \lambda_{\alpha}^{-1} \partial_x S^x +  (2\lambda_{\alpha}^{-2} + \lambda_{\rm{ext}}^{-2}) S^z \label{eq_sde_1z}\; ,\\
   &&\partial^2_x S^y = (\lambda_{\alpha}^{-2} + \lambda_{\rm{ext}}^{-2}) S^y + \frac{\gamma\sigma_{\rm{D}}}{D} \lambda_{\alpha}^{-3} E_x \label{eq_sde_1y}\; .
\end{eqnarray}
Notice that for generality we assume different ESR lengths in the Rashba and in the normal conductor, $\lambda_{\rm{ext}}$ and $\lambda_{\rm{s}}$, respectively.   

We solve Eqs.~\eqref{eq_diff_N} and~\eqref{eq_sde_1x}--\eqref{eq_sde_1y} in two different situations. We first consider the inverse SGE [Fig.~\ref{fig_sketch}(c)]: a finite spin density is induced in the Rashba conductor due to the applied  electric field. We explore whether such spin density can diffuse into the normal region. We then focus on the reciprocal situation [Fig.~\ref{fig_sketch}(d)] in which we assume that a spin density is created (e.g., by injection) at some point at the normal conductor and determine the charge current  induced at the Rashba conductor via the SGE. 

\section{Charge-to-spin conversion: the inverse spin-galvanic effect}
\label{sec_cs}
In this section we explore  the  charge-to-spin conversion in homogeneous and hybrid systems with intrinsic SOC. 
This  conversion leads to the inverse SGE, which in the particular case of  Rashba SOC, is also called the Edelstein effect~\cite{Edelstein1990}.  

We  start our discussion by analyzing this effect in a bulk material  with  intrinsic SOC. Even though  this example has been widely studied in the literature~\cite{sinova2004universal,mishchenko2004spin,raimondi2005spin}, its  discussion here will serve as good starting point to introduce the main physical parameters used in the subsequent analysis of a more complicated hybrid setup.    

\subsection{Homogeneous material with intrinsic SOC}
\label{sec_bulk}
The question under which conditions  a charge current through  a conductor  with intrinsic linear-in-momentum SOC can create a transverse SH current, was addressed in several works  (see, e.g., Refs.~\cite{sinova2004universal,mishchenko2004spin,raimondi2005spin}).  Here, we show  how the answer to this question can be found straightforwardly from the SU(2)-covariant spin diffusion equation. 

In a bulk  homogeneous system  the spin density has no spatial dependence  and,  therefore, the diffusion equation reduces to an algebraic equation  after setting  the spatial derivatives in Eq.~\eqref{eq_diff_charge0} to zero. In the presence of an external electric field, this equation reduces to:
\begin{equation}
    \label{eq:EEspin_bulk}
    - \Gamma^{ab} S^b = \gamma\sigma_{\rm{D}} \mathcal{A}_k^b \mathcal{F}_{ki}^c \varepsilon^{abc} E_i + \frac{1}{\tau_{\rm{ext}}} S^a\; .
    \end{equation}
We first assume that $\tau_{\rm{ext}}^{-1}=0$, and obtain: 
\begin{equation}
S^a_{\rm{int}} = \frac{\gamma\sigma_{\rm{D}}}{D} \mathcal{A}_i^a E_i\; .
\end{equation} 
The SH current, i.e., the spin current transverse to the applied electric field, is obtained from  Eq.~\eqref{eq:gen_spin_current} and reads as
\begin{equation}
j_k^a =  - D \mathcal{A}_k^c \varepsilon^{cba} \left( S^b - S^b_{\rm{int}} \right)\; .
\label{eq_sHc_0}
\end{equation}
In the absence of ESR, $\hat{S}=\hat{S}_{\rm{int}}$, and hence   
no transverse current is generated. The spin current induced by the SU(2) magnetic field is fully compensated  by the spin diffusive current. This also means that in a homogeneous finite system with intrinsic SOC, no spin accumulation at the boundary is expected.  This can be seen as a direct consequence of the SU(2)-covariant conservation of the current, Eq.~\eqref{eq_cont}, which has to be zero at the sample boundaries.

The situation is quite different in the presence of a finite ESR.  For the Rashba SOC the solution of Eq.~\eqref{eq:EEspin_bulk} can be explicitly written:
\begin{equation}
\label{eq_EE_pure2}
S^a = \frac{S^a_{\rm{EE}}}{1+r^2_{\rm{ext}}}\; ,
\end{equation}
where the parameter $r_{\rm{ext}} = \lambda_{\alpha} / \lambda_{\rm{ext}}$, with  $\lambda_{\rm{ext}} = \sqrt{\tau_{\rm{ext}} D}$, characterizes the relative strength of ESR, and 
\begin{equation}
\label{eq_EE_pure}
S^a_{\rm{EE}} = \frac{\gamma\sigma_{\rm{D}}}{D\lambda_\alpha}  \epsilon^{zai} E_i
\end{equation} 
is the well-known Edelstein result~\cite{Edelstein1990, Lyanda-Geller89} for the current-induced spin in a Rashba conductor~\footnote{ Notice that we distinguish between $S_{\rm{int}}$ and $S_{\rm{EE}}$, to emphasize that the latter is valid specifically for Rashba SOC. In contrast,  $S_{\rm{int}}$ denotes the spin density induced by the electric field for any linear-in-momentum SOC.}. 

Substitution of Eq.~\eqref{eq_EE_pure} into  Eq.~\eqref{eq_sHc_0}, leads to a   finite SH current~\cite{ chalaev2005spin, gorini2008spin, raimondi2012spin}: 
\begin{equation}
    j_k^z =  \frac{D} {\lambda_{\alpha}} \frac{r^2_{\rm{ext}}}{1+r^2_{\rm{ext}}}  S^k_{\rm{EE}}\; .
\label{eq_sHc}
\end{equation}
The above results are  used in the next sections  to contrast them with those obtained  for  hybrid systems. 

\subsection{Hybrid structure with inhomogeneous SOC}
\label{sec_inv_SGE}
We now focus on the hybrid diffusive structure sketched in Fig.~\ref{fig_sketch}(c). The structure can be viewed as a building block of a lateral spin valve [Fig.~\ref{fig_sketch}(b)], commonly used for  non-local detection of magnetoelectric effects~\cite{valenzuela2006direct,villamor2015modulation}.
The charge-to-spin conversion can be detected by passing a charge current at the Rashba conductor [blue region in Fig.~\ref{fig_sketch}(b)]. 
This current generates a  spin accumulation which can diffuse into the normal conductor (gray region) and can be detected 
as a spin voltage measured by a  ferromagnetic electrode (orange region)~\cite{jedema2001electrical}.

In our model of Fig.~\ref{fig_sketch}(c), the normal conductor occupies the half-plane $x < 0$ and the Rashba conductor is at $x > 0$. We  solve the diffusion equations in both regions [Eqs.~\eqref{eq_diff_N} and~\eqref{eq_sde_1x}--\eqref{eq_sde_1y}] together with the BC at $x=0$ [Eqs.~\eqref{eq_bc_j} and~\eqref{eq_bc_s}]. 

In the normal conductor region, $x<0$, the solution of  Eq.~\eqref{eq_diff_N} is an exponential  function decaying away from the interface over the spin diffusion length $\lambda_{\rm{s}}$.  Whereas, the solution at the Rashba conductor depends on the direction of the applied electric field.  We distinguish between two different situations: when the electric field is applied  parallel or perpendicular to the interface. 
\begin{figure}
    \centering
    \includegraphics[width=1\columnwidth]{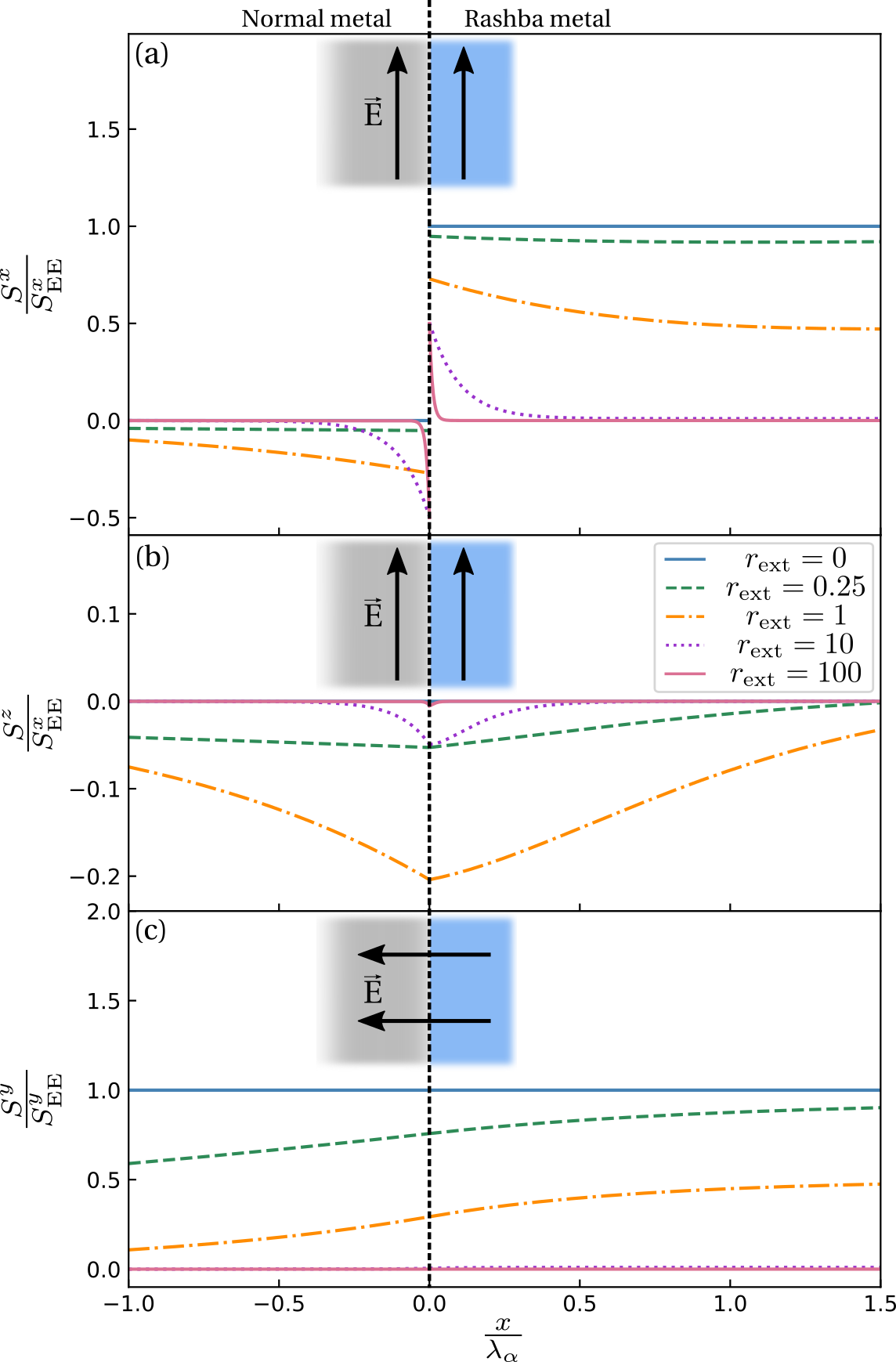}
    \captionsetup{justification= raggedright}
        \caption{Spatial dependence of the spin density induced by applying an electric field [see Fig.~\ref{fig_sketch}(c)] for different values of $r_{\rm{ext}}$. We distinguish two possible directions of the electric field: (a), (b) parallel, and (c) perpendicular to the interface. In all figures it is assumed that $\lambda_{\rm{s}} = \lambda_{\rm{ext}}$, and the calculated  spin density is normalized by the corresponding bulk value  $S^a_{\rm{EE}}$ of Eq.~\eqref{eq_EE_pure}.}
    \label{fig_S_NR}
\end{figure}

\subsubsection{Electric  field  parallel to the interface: ${\bf{E}}=E_y \hat{\bf e}_y$} \label{subsec_Epa}
If  the electric field  is applied parallel to the interface between the normal and Rashba conductors then, according to  Eq.~\eqref{eq_EE_pure},   the induced spin density in the bulk of the Rashba conductor is polarized perpendicular to ${\bf{E}}$, which in our case corresponds to  the direction $S^x$.
From Eqs.~\eqref{eq_sde_1x} and~\eqref{eq_sde_1z} we see that the diffusion of this component is coupled to $S^z$, whereas the spin polarization in the $y$ direction is not induced.   
Thus, one needs to solve two coupled linear second-order  differential equations with BC at the interface between the normal and the Rashba conductors obtained from   Eqs.~~\eqref{eq_bc_j} and~\eqref{eq_bc_s}: 
\begin{equation}
\label{eq_bc_Epa}
j_x^a|_{0^+}\ = j_x^a|_{0^-}\; , \ \ \ S^a |_{0^+} - S^a |_{0^-} = S_{\rm{EE}}^x\delta^{ax}\; .
\end{equation}
The explicit form of the spatial dependence of $S^x$ and $S^z$ is given in Eq.~\eqref{eq_app_Epa} of Appendix~\ref{app_cs} and it is 
shown in Figs.~\ref{fig_S_NR} (a) and (b), respectively, for $\lambda_{\rm{ext}} = \lambda_{\rm{s}}$. 
The obtained behavior can be easily understood from the bulk solution. When the ESR is negligibly small, $r_{\rm{ext}}\rightarrow 0$, the SH current is zero and the Edelstein spin density cannot diffuse into the normal conductor, solid blue line in Fig.~\ref{fig_S_NR}(a). This is a consequence of the SU(2)-covariant conservation of the spin. Such conservation does not hold   for a finite $r_{\rm{ext}}$. Indeed, ESR  leads to a finite spin current in the Rashba conductor, Eq.~\eqref{eq_sHc}, and  consequently,  the continuity of the spin current at the interface leads to a diffusive spin current  in the normal conductor. This mechanisms of spin injection into the normal conductor is different from the one discussed in Ref.~\cite{brataas2007spin}, in which the spin injection takes place due to different momentum relaxation time at both sides of the interface.

As mentioned above, the precession terms in Eqs.~\eqref{eq_sde_1x} and \eqref{eq_sde_1z} couple the $S^x$ and $S^z$ components and therefore both are induced in the whole system, as shown in Fig.~\ref{fig_S_NR}(b).  Far away from the interface inside the Rashba conductor,  $x/\lambda_{\alpha} \gg 1$, the spin density   reaches its  bulk value given by  Eq.~\eqref{eq_EE_pure2}.

One can obtain simple expressions for the spatial dependence of the spin density  in limiting cases.  
For example, if the ESR is very small, $r_{\rm{ext}} \ll 1$, we obtain from Eq.~\ref{eq_app_Epa}:
\begin{equation}
\begin{split}
&\frac{S^x(x)}{S_{\rm{EE}}^x} \approx  \Theta(x) - \Theta(-x)  r^2_{\rm{ext}} e^{\frac{x}{\lambda_{\rm{s}}}}\;  , \\
&\frac{S^z(x)}{S_{\rm{EE}}^x} \approx - r_{\rm{ext}}^2 \left( \Theta(x) \operatorname{\mathbb{I}m} \left\{ \frac{\kappa_0^{*2}}{2\sqrt{2}} e^{-\frac{\kappa_0 x}{\lambda_{\alpha}}} \right\} + \Theta(-x) e^{\frac{x}{\lambda_{\rm{s}}}} \right)\; , \label{eq:sxz_para}\\
\end{split}
\end{equation}
where $\kappa_0^2 = (- 1 + i\sqrt{7})/2$, and corresponds to the definition of $\kappa$ in Eq.~\eqref{eq_a_kappa} with $r_{\rm{ext}} = 0$.
This means that, to leading order in $r_{\rm{ext}}$, the  Edelstein spin given by Eq.~\eqref{eq_EE_pure2} is induced homogeneously in the Rashba conductor, whereas the amplitude of the spin density that  diffuses into  the normal region  is  proportional to $r_{\rm{ext}}^2$ [cf. Figs.~\ref{fig_S_NR}(a) and (b)]. When  $r_{\rm{ext}} = 0$, we recover the results obtained in Refs.~\cite{tserkovnyak2007boundary, raimondi2006quasiclassical}.

In the opposite limit, i.e., when  $r_{\rm{ext}}\gg1$, we obtain from Eq.~\eqref{eq_app_Epa}: 
\begin{equation}
\label{eq_limitgg}
    \begin{split}
        &\frac{S^x (x)}{S_{\rm{EE}}^x} \approx \frac{\lambda_{\rm{s}}}{\lambda_{\rm{s}} + \lambda_{\rm{ext}}} \left( \Theta(x) \frac{\lambda_{\rm{ext}}}{\lambda_{\rm{s}}} e^{-\frac{x}{\lambda_{\rm{ext}}}} - \Theta(-x) e^{\frac{x}{\lambda_{\rm{s}}}} \right)\; , \\
        &\frac{S^z (x)}{S_{\rm{EE}}^x} \approx \frac{- 1}{r_{\rm{ext}}}\frac{ 1 + \frac{\lambda_{\rm{ext}}}{\lambda_{\rm{s}}}}{ \frac{\lambda_{\rm{ext}}^2}{\lambda_{\rm{s}}^2} + 2\frac{\lambda_{\rm{ext}}}{\lambda_{\rm{s}}} + 1} \left( \Theta(x) e^{-\frac{x}{\lambda_{\rm{ext}}}} + \Theta(-x) e^{\frac{x}{\lambda_{\rm{s}}}} \right)\; . \\
    \end{split}
\end{equation}
In this case, the induced $S^x$ is localized  at the interface and decays exponentially into both conductors [cf.  Fig.~\ref{fig_S_NR}(a)].  The sign of the spin at both sides of the interface is opposite. If $\lambda_{\rm{ext}}=\lambda_{\rm{s}}$, as in Fig.~\ref{fig_S_NR}, the value of the spin at each side of the  interface is  $\pm S^x_{\rm{EE}}/2$. Due to the Rashba coupling, a small contribution polarized in $z$ also appears as shown in Fig.~\ref{fig_S_NR}(b). If $\lambda_{\rm{ext}}\neq\lambda_{\rm{s}}$, we distinguish two cases:
If  $\lambda_{\rm{s}} \ll \lambda_{\rm{ext}}$, then the spin relaxes strongly  in the normal conductor next to the interface. On the Rashba side, $S^z$ practically disappears whereas there is an $x$-polarized spin accumulation at the edge  of the order of $S_{\rm{EE}}^x$,  which decays towards the bulk.
In  the opposite limit,  $\lambda_{\rm{s}} \gg \lambda_{\rm{ext}}$, the spin density in the Rashba conductor is strongly suppressed  by the ESR. Whereas, at the interface  in the normal conductor side 
 a spin density  $S^x_{\rm{EE}}$ appears and decays   over $\lambda_{\rm{s}}$ into  the normal conductor.
\begin{figure}
    \centering
    \includegraphics[width=1\columnwidth]{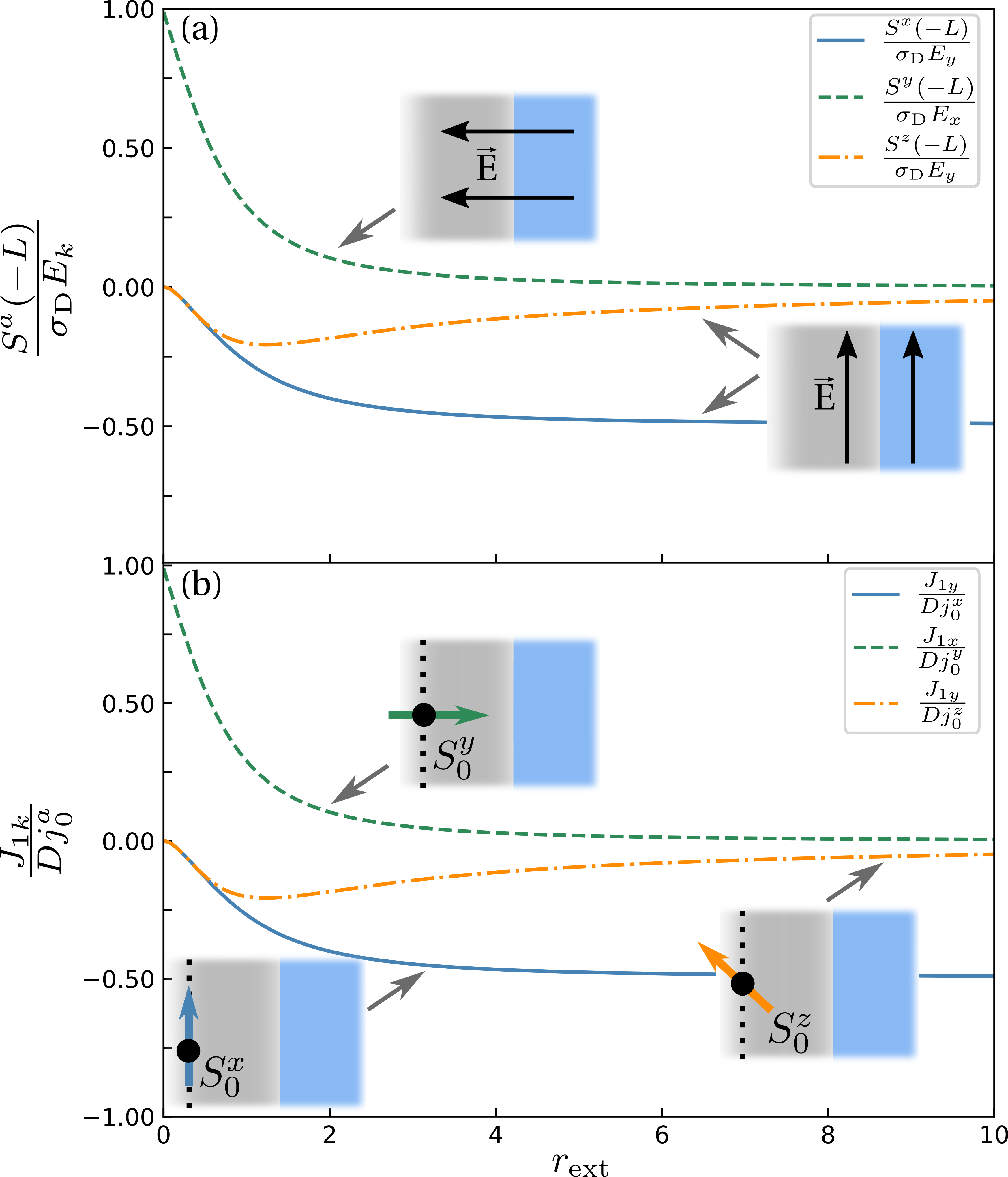}
    \captionsetup{justification= raggedright}
    \caption{(a) Spin density induced at $x=-L$ by an electric field [see Fig.~\ref{fig_sketch}(c)] as a function of $r_{\rm{ext}}$. We distinguish between an electric field applied parallel and perpendicular to the interface. (b) Integrated charge current induced by a spin density injection at $x=-L$ [see Fig.~\ref{fig_sketch}(d)] as a function of $r_{\rm{ext}}$. The charge current flows in different directions depending on the polarization of the injected spin density.  All curves are calculated in units of $\gamma/(D\lambda_{\alpha})$, for $L/\lambda_{\rm{s}} = 0.01$, $\lambda_{\rm{s}} = \lambda_{\rm{ext}}$, and normalized according to Eq.~\eqref{eq_rec}.}
    \label{fig_SL_Jy}
\end{figure}

The spin density induced in the normal conductor can be measured 
by detecting a spin voltage with a local ferromagnetic probe. We 
assume that such a contact is  located at a distance $L$ from the interface [see Figs.~\ref{fig_sketch}(b) and \ref{fig_sketch}(c)]. 
In Fig.~\ref{fig_SL_Jy}(a),  we show the dependence on  $r_{\rm{ext}}$ of both spin components (solid blue and dashed-dotted orange lines) induced in the normal conductor [Eq.~\eqref{eq_app_Epa}] at the detector. 
We chose $\lambda_{\rm{s}} = \lambda_{\rm{ext}}$ and $L\ll\lambda_{\rm{s}}$.
As explained above, in the absence of ESR, $r_{\rm{ext}} = 0$, the spin density induced by the charge current in  the Rashba conductor does not diffuse into the normal part and hence both components are zero. 
For finite ESR, both $S^{x}$ and $S^{z}$ become finite at $x=-L$, but their dependence on $r_{\rm{ext}}$ is quite different. The absolute value of $S^x$ increases  monotonically with $r_{\rm{ext}}$ and asymptotically approaches $S^x_{\rm{EE}}/2$, while $S^z$ reaches a maximum at $r_{\rm{ext}}\approx1$ and  decays towards zero by further increase of $r_{\rm{ext}}$.

It is worth noticing that the situation of a parallel electric field explored in this section also corresponds to the experimental situation of  Ref.~\cite{kim2017evaluation}: a normal metal film, Cu, is in contact with  an  insulator, Bi$_2$O$_3$ forming what the authors of that work  called  a Rashba interface. This setup is sketched  in Fig.~\ref{fig_sketch}(a). If one assumes that the SOC is confined to the blue layer,  and the electric field is applied along the films, then,  in the absence of ESR no  spin current may flows  from the interface into the Cu layer.   Therefore, the observed magnetoresistance associated  to a finite spin current cannot be only due to the Edelstein effect at the interface, but can be attributed to other extrinsic spin relaxation sources,  as predicted from our previous  analysis.

\subsubsection{Electric field perpendicular to the interface: $ {\bf{E}}=E_x \hat{\bf e}_x$} \label{subsec_Eperp}
Now we focus on the  situation in which the electric field is applied perpendicular to the interface.  According to Eq.~\eqref{eq_EE_pure}, the induced spin density is polarized in the direction of ${\hat{\cal{A}}}_x$, which for Rashba SOC corresponds to $S^y$.  This component is decoupled from the other two, see Eqs.~\eqref{eq_sde_1x}-\eqref{eq_sde_1y}, and therefore in this case  $S^{x,z}=0$.   

For the perpendicular orientation of the electric field, the BC correspond to the continuity of both the spin current and the spin density, Eqs.~\eqref{eq_bc_j} and~\eqref{eq_bc_s}:
\begin{equation}
\label{eq_bc_Epe}
j_x^y|_{0^+}\ = j_x^y|_{0^-}\; , \ \ \ S^y |_{0^+} = S^y |_{0^-}\; .
\end{equation}
This means that the spin generated at the Rashba conductor via the Edelstein effect, diffuses into the normal conductor, even in  the  absence  of  any ESR mechanism.
From Eqs.~\eqref{eq_diff_N},~\eqref{eq_sde_1y}, and~\eqref{eq_bc_Epe}, one can determine  explicitly the spatial dependence of  $S^y$. In the normal conductor it reads ($x<0$):
\begin{equation}
\label{Sy_NM}
\begin{array}{c}
{S^y (x)} =  \frac{{S_{\rm{EE}}^y}}{\sqrt{1 + r_{\rm{ext}}^2}} \frac{\lambda_{s} }{\lambda_{\alpha} + \lambda_{s}\sqrt{1 + r_{\rm{ext}}^2}}e^{\frac{x}{\lambda_{\rm{s}}}}\; ,
\end{array}
\end{equation}
and in the Rashba region ($x>0$):
\begin{equation}
\label{Sy_RM}
\begin{array}{c}
{S^y (x)} = \frac{S_{\rm{EE}}^y}{1+r_{\rm{ext}}^2} \left(1- \frac{\lambda_{\alpha}}{\lambda_{\alpha}+\lambda_s\sqrt{1+r_{\rm{ext}}^2}} e^{-\frac{x}{\lambda_{\alpha}} \sqrt{1+r_{\rm{ext}}^2}}  \right) \; .
\end{array}
\end{equation}
This result has to be contrasted to the one obtained when the field is applied parallel to the interface.  {Namely,  in the latter case when $r_{\rm{ext}} = 0$ no spin diffuses into the normal conductor.   Here, however,  even if $r_{\rm{ext}}=0$, the diffusion occurs as a consequence of the broken translation symmetry in the direction of the electric field. }

In Fig.~\ref{fig_S_NR}(c) we show the spatial dependence of $S^y$, assuming that the ESR in the normal and Rashba conductors are equal, $\lambda_{\rm{s}} = \lambda_{\rm{ext}}$.  As  in Figs.~\ref{fig_S_NR}(a) and (b), deep in the Rashba conductor, $\frac{x}{\lambda_{\alpha}} \gg 1$, one obtains the  bulk value for the spin density, determined by   Eq.~\eqref{eq_EE_pure2}. 
Because of the continuity of the spin density, the increase of  ESR  leads to an overall homogeneous decrease of the spin density. 

We compute the measurable spin density at a distance $L$ from the interface, see Figs.~\ref{fig_sketch}(b) and (c).  
It is shown  in Fig.~\ref{fig_SL_Jy}(a) (dashed green line) for the particular case of $\lambda_{\rm{s}} = \lambda_{\rm{ext}}$ and $L \ll \lambda_{\rm{s}}$. Due to the latter condition, for  $r_{\rm{ext}} = 0$  the spin density at $x=-L$ is approximately equal  to  $S^y_{\rm{EE}}$. When the ESR is switched on, the current induced spin in the bulk Rashba conductor decreases monotonically according to Eq.~\eqref{eq_EE_pure2}, and so does the spin density value at $x=-L$. 

The spin generated in the normal conductor is associated to a diffusive spin current  $j^y_x$, parallel to the electric field as a consequence of the spatial variation of $S^y$, Eqs.~\eqref{Sy_NM} and~\eqref{Sy_RM}.

But more interesting is the appearance of a SH current, $j_y^z$, in the Rashba conductor as a consequence of both, the covariant diffusion  and  the SU(2) magnetic field. 
 This is a transverse to the electric field current  and  it can be calculated from  Eq.~\eqref{eq:gen_spin_current}:
\begin{equation}
{j}_y^z(x)= -\frac{D}{\lambda_\alpha}\left(S^y(x)-S_{\rm{EE}}^y\right)\; , 
    \label{eq_SHE_perp}
\end{equation}
which after substitution of  $S^y(x)$ from  Eq.~\eqref{Sy_RM}, gives:
\begin{equation}
{j}_y^z(x) = D S^y_{\rm{EE}} \left( \frac{r_{\rm{ext}}^2}{\lambda_\alpha(1+r_{\rm{ext}}^2)}+\frac{ e^{-\frac{x}{\lambda_{\alpha}} \sqrt{1+r_{\rm{ext}}^2}}}{\lambda_{\alpha}+\lambda_s\sqrt{1+r_{\rm{ext}}^2}} \right)\; .
\end{equation}
The first term is the  bulk solution of Eq.~\eqref{eq_sHc}, whereas the second term is  a correction due to the broken translation symmetry in the direction of the field. Interestingly, even in the absence of ESR, $r_{\rm{ext}}=0$,  there is a finite contribution to the  SH  current which is maximized  at the interface and decays  exponentially into the bulk. Such a localized SH current resembles the one obtained in Ref.~\cite{mishchenko2004spin}  in  a different geometry and for $r_{\rm{ext}}=0$.

\section{Reciprocity between the non-local spin-galvanic effect and its inverse}
\label{sec_rec}
In the previous section we discuss  the  non-local inverse SGE:  a finite  spin density,  detectable in the normal conductor at a distance $L$ from the interface [Figs.~\ref{fig_sketch}(b) and (c)] is induced as a response to an electric field applied  both parallel and perpendicular to the interface.   
In the next section, we explore the reciprocal non-local effect, i.e.,  the charge current induced by a spin injection into the normal conductor [gray region in Figs.~\ref{fig_sketch}(b) and (d)].
Before analyzing this effect for this specific  geometry, we  examine  the diffusion equation  and  identify  a general non-local reciprocity between the  spin induced by a charge current and the  spatially  integrated charge current induced by spin injection.  We interpret this reciprocity as the non-local version of the reciprocity between the SGE and its inverse. 

Our starting point is the general spin diffusion equation, Eq.~\eqref{eq_diff_charge0}, that we rewrite as follows:
\begin{equation}
\label{eq_deq_G}
 \left( D\tilde\partial_k\tilde\partial_k - \tau_{\rm{ext}}^{-1} \right) \hat S = \gamma \sigma_{\rm{D}} \tilde{\partial}_i \hat{\mathcal{F}}_{ik} E_k\; .
\end{equation}
We assume, as before,  that the SOC is inhomogeneous with a spatial variation over lengths larger than the mean-free path.  As shown in Sec.~\ref{sec_model}, the BC for  hybrid interfaces can be obtained by integration of this equation. Here, instead, we keep the spatial dependence in the SU(2) fields and work with the general Eq.~\eqref{eq_deq_G}.  
We assume that the fields only vary in one direction, which we define as $x$. This is our only assumption.  Thus, the diffusion equation reduces to a 1D linear differential equation. 
The solution can be written as follows: 
\begin{equation}
\label{eq_S_G}
    \hat S(x)  = \gamma \sigma_{\rm{D}} E_k \int_{-\infty}^{\infty} \hat{G}(x, x') \tilde{\partial}_i \hat{\mathcal{F}}_{ik}(x')\ dx'\; ,
\end{equation}
where the Green's function $\hat G$ satisfies: 
\begin{equation}
\label{diff_GF}
\left( D\tilde\partial_k\tilde\partial_k - \tau_{\rm{ext}}^{-1} \right)\hat{G}(x, x')=\delta(x-x')\; .
\end{equation}
Equation~\eqref{eq_S_G} describes the non-local inverse SGE, i.e., the spin density created at $x$ by an homogeneous electric field in the $k$ direction.

We now consider  the spin-to-charge conversion described by Eq.~\eqref{eq:charge_current}, which can be rewritten as:
\begin{equation}
\label{eq_jk_jk1}
j_k = -\sigma_{\rm{D}} \partial_k \mu + j_{1k} \; ,
\end{equation}
where $\mu$ is the electrochemical potential defined by:
\begin{equation}
\sigma_{\rm{D}}\partial_k\mu= D \partial_kn + \sigma_{\rm{D}}E_k\; ,
\label{eq_mu}
\end{equation}
and:
\begin{equation}
    j_{1k}= -\gamma \hat{\mathcal{F}}_{ik} \hat{j}_i \; ,
    \label{eq_an_current}
\end{equation}
is the charge current density generated via the SOC. 
Here, $j_i^a$ is the spin current originated from the covariant diffusion of $S^a$ in the $i$ direction and described by the first term in Eq.~\eqref{eq:gen_spin_current}. 
We now integrate Eq.~\eqref{eq_an_current} over $x$ and obtain:
\begin{equation}
\label{eq_J_G}
    J_{1k} = \gamma D \int_{-\infty}^{\infty} \hat{\mathcal{F}}_{ki} \tilde\partial_i \hat S \ dx = \gamma D \int_{-\infty}^{\infty} \hat S \tilde\partial_i \hat{\mathcal{F}}_{ik} \ dx\; ,
\end{equation}
where the last equality follows from  integration  by parts. 
Equation~\eqref{eq_J_G} describes the charge current density generated by a spin density $\hat S$ and integrated over the direction of the spatial inhomogeneity. The spin density  has to be calculated  from the diffusion equation.  Here, we assume that no electric field is applied, but instead a spin current $\hat{j}_0$ is injected locally at $x=x_0$. The diffusion equation~\eqref{eq_deq_G}, corresponding to this situation, reads as 
\begin{equation}
\label{eq_BC_j0}
 \left( D\tilde\partial_k\tilde\partial_k - \tau_{\rm{ext}}^{-1} \right) \hat S  = \hat{j}_0 \delta(x-x_0)\; .
\end{equation}
The solution to this equation can be written again in terms of the Green's function [Eq.~\eqref{diff_GF}] as follows:
\begin{equation}
    \hat S(x) = \hat{j}_0 \hat{G} (x_0, x)\; .
\end{equation}
Substituting this result  into Eq.~\eqref{eq_J_G}, and comparing it with Eq.~\eqref{eq_S_G}, leads to the following  relation between the $k$ component of the induced charge current and the $a$ component  of the induced  spin density:
\begin{equation}
\label{eq_rec}
    \frac{J_{1k}}{Dj_0^a} = \frac{S^a(x_0)}{\sigma_{\rm{D}} E_k}\; .
\end{equation}
That is a remarkable result that connects the integrated charge current induced by a spin current injected at $x=x_0$, with the spin density at $x_0$ induced by an applied electric field. This non-local  reciprocity is a general property for any diffusive system with a 1D spatial inhomogeneity. 
It explains the identical curves shown in both panels of  Fig.~\ref{fig_SL_Jy}. Specifically, the result of Fig.~\ref{fig_SL_Jy}(b) is discussed in the next section.
\section{Spin-to-charge conversion: the spin-galvanic effect}
\label{sec_sc}
We now verify  the reciprocity demonstrated in the previous section by computing explicitly the non-local SGE  in the setup sketched in Fig.~\ref{fig_sketch}(d). We  assume that a spin current $\hat{j}_0$  is injected into the normal conductor at $x=-L$. Experimentally this can be done, for example, by injecting a  current  from a ferromagnetic lead~\cite{jedema2001electrical,jedema2002electrical}. 
We first solve the spin diffusion equations~\eqref{eq_diff_N} and~\eqref{eq_sde_1x}--\eqref{eq_sde_1y}
together with the BC of Eqs.~\eqref{eq_bc_j} and~\eqref{eq_bc_s}. Since ${\bf E}=0$,  the latter imply the conservation of the spin current and the spin density at the interface located at $x=0$:
\begin{equation}
\label{eq_bc_Sinj}
j_x^a|_{0^+}\ = j_x^a|_{0^-}\; , \ \ \ S^a |_{0^+} = S^a |_{0^-}\; .
\end{equation}
This continuity leads to the spin diffusion into the Rashba conductor for any polarization of the injected spin and strength of the ESR. At the injection point, $x=-L$, the continuity of the spin density is assumed and, from Eq.~\eqref{eq_BC_j0},
\begin{equation}
\label{eq_bc_j0}
    j_x^a|_{-L^+}\ - j_x^a|_{-L^-} = j_0^a\; ,
\end{equation}
where  $j_0^a$ is the injected spin current. Again, in Eqs.~\eqref{eq_sde_1x}--\eqref{eq_sde_1y} we see that  
the components $S^{x}$ and $S^z$ are coupled through the SOC whereas the $S^y$ component is not. 
Therefore, in the next two subsections we distinguish between the injected spin current polarized in the $x$ and $z$ directions, and the injected current polarized in the $y$ direction. As shown in Sec.~\ref{sec_rec}, these two cases  should be reciprocal to the  results of Sec.~\ref{sec_cs}  when the  electric field was applied parallel or perpendicular to the interface, respectively.
\subsection{Spin current polarized in the $x$ or  $z$ direction}
\label{subsec_Rcoupled}
\begin{figure}
    \centering
    \includegraphics[width=1\columnwidth]{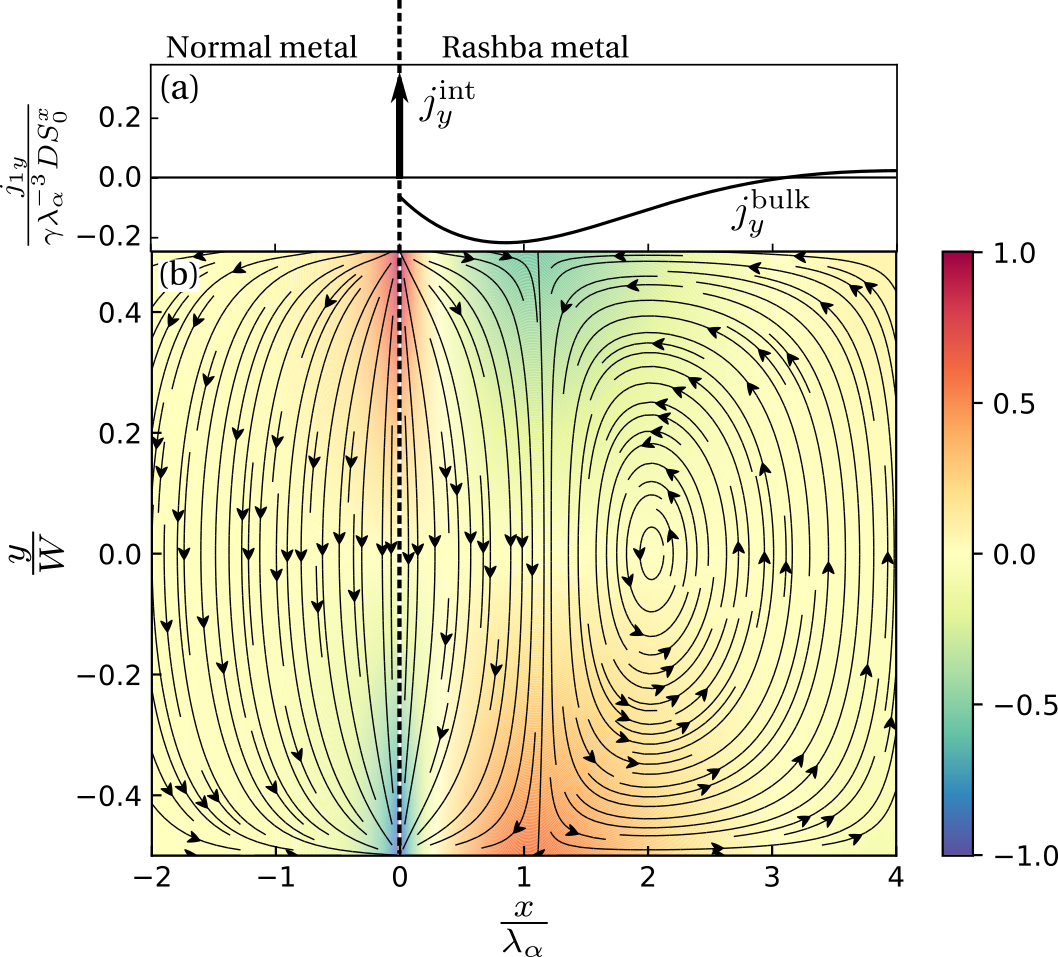}
    \captionsetup{justification= raggedright}
    \caption{(a) Spatial dependence of the charge current density $j_{1y}$ induced by the Rashba SOC when the spin density injected at $x=-L$ is $x$-polarized. The interfacial and bulk contributions can be distinguished. (b) Redistribution of the electrochemical potential $\mu$ due to the insulating boundaries placed at $y=\pm W/2$. The vector field lines correspond to the total charge current densities, ${\bf{j}} = - \sigma_{\rm{D}} \nabla \mu + j_{1y} {\bf{\hat{e}}}_y$. Both plots are shown for pure Rashba SOC, $r_{\rm{ext}}=0$, with $\lambda_{\alpha}/\lambda_{\rm{s}} = 0.1$, and $L/\lambda_{\rm{s}} = 0.1$}
    \label{fig_red_mu}
\end{figure}

Let us assume that the spin current injected at $x=-L$ 
is polarized in the $x$ or $z$ direction and compute the charge current density  induced in the Rashba conductor [Eq.~\eqref{eq_an_current}]. Since the Rashba SOC is only finite at $x>0$, this current flows in the $y$ direction and  consists of two contributions:
\begin{equation}
\label{eq_anom_cc1}
j_{1y}(x) = j_y^{\rm{int}} + j_y^{\rm{bulk}} = \frac{\gamma}{ \lambda_{\alpha}} \Big( \delta(x) j_x^x(x) + \Theta(x)  \lambda_{\alpha}^{-1} j_x^z(x) \Big)\; ,
\end{equation}
with:
\begin{equation}
 j_x^x (x)= - D \left( \partial_x S^x - \frac{S^z}{ \lambda_{\alpha}}\right)\; , \ \ \operatorname{and}\ \ \ 
 j_x^{z} (x)= - D \left(\partial_x S^z + \frac{S^x}{ \lambda_{\alpha}} \right)\; ,
\end{equation}
from Eqs.~\eqref{eq_jx_E1} and \eqref{eq_jx_E3}. The explicit spatial dependence of $S^x$ and $S^z$ is given  in Eq.~\eqref{eq_app_S0xz} of Appendix~\ref{app_sc}.
One of the contributions, $ j_y^{\rm{int}}$,  is localized at the interface, whereas the other one, $ j_y^{\rm{bulk}}$, is finite at the Rashba conductor. In  Fig.~\ref{fig_red_mu}(a) we show the spatial dependence of the latter  in the absence of ESR,  $r_{\rm{ext}} = 0$. In view of the result of Sec.~\ref{subsec_Epa}, at a first glance it might seem strange that, even though $r_{\rm{ext}} = 0$, a finite charge current density is induced  in the system.  
However, as we have understood in the previous section with Eq.~\eqref{eq_rec}, the reciprocity  involves the integrated  current. Indeed, by substituting Eq.~\eqref{eq_app_S0xz} into Eq.~\eqref{eq_anom_cc1}, and performing the integration, one can demonstrate indeed  that  $J_{1y}=0$ if $r_{\rm{ext}} = 0$.  

In contrast, in the case of a finite $r_{\rm{ext}}$ one obtains a finite total current (see Appendix~\ref{app_Jk}): 
\begin{equation}
\label{eq_Jy}
 J_{1y} = \gamma \lambda_{\alpha}^{-1} \int_{0}^{\infty} \frac{S^x}{\tau_{\rm{ext}}} dx\; .
\end{equation}
This result is  in accordance with the reciprocity relation [Eq.~\eqref{eq_rec}] and the results of Sec.~\ref{subsec_Epa}. 
In Fig.~\ref{fig_SL_Jy}(b) we show the behavior of $J_{1y}$ (solid blue and dashed-dotted orange lines) as a function of $r_{\rm{ext}} $. 
With the proper normalization imposed by Eq.~\eqref{eq_rec}, these curves are identical to those in Fig.~\ref{fig_SL_Jy}(a). 

\subsection{Spin current polarized in the $y$ direction}
\label{subsec_Runcoupled}
If the spin current  injected at $x=-L$ is $y$-polarized,  only the   $S^y$ component  is induced.  This implies that only the longitudinal charge current density with $k=x$ in Eq.~\eqref{eq_an_current} is non-zero:
\begin{equation}
\label{eq_j_1x}
j_{1x} = -\Theta(x)\gamma\lambda_{\alpha}^{-2}j_y^z= \Theta(x)D \gamma\lambda_{\alpha}^{-3}S^y\; ,
\end{equation}
where we have used Eq.~\eqref{eq:gen_spin_current} for the spin current density. In the present geometry, only the commutator part of the covariant derivative contributes, such that 
 the current $j_y^z$ is  proportional to $S^y$.
The analytic expression for   $S^y(x)$ is given in Eq.~\eqref{eq_app_S0y} of Appendix~\ref{app_sc}. {In the  region $x>0$ it reads as
\begin{equation}
 \label{Sy}
 S^y(x) = \frac{\lambda_{\alpha} j_0^y}{D} \frac{e^{- \left( \sqrt{1+r_{\rm{ext}}^2} \frac{x}{\lambda_{\alpha}} + \frac{L}{\lambda_{\rm{s}}} \right)}}{r_{\rm{s}} + \sqrt{1+r_{\rm{ext}}^2}}\; .
\end{equation}
According to  Eq.~\eqref{eq_jk_jk1},  the charge current density is given by the contribution $j_{1x}$ and the diffusive term:}
\begin{equation}
 \label{j-total}
 j_x = -\sigma_{\rm{D}}\partial_x\mu(x) + j_{1x}(x)\; .
\end{equation}
Because of the charge conservation, $\partial_x j_x = 0$ and the total charge current should be constant in space, $j_x=$~const. The value of this constant is determined by the BC imposed on the outer boundaries of the system. For example, in a large, but finite, sample with floating edges, $j_x=0$. This condition, together with Eq.~\eqref{j-total}, determines the distribution of the electrochemical potential $\mu(x)$ in the system:
\begin{equation}
\partial_x\mu(x) = \frac{1}{\sigma_{\rm{D}}} j_{1x}(x)\; ,    
\end{equation}
and eventually relates the voltage drop $\Delta\mu$ across the sample to the space integrated induced current:
\begin{equation}
 \label{Delta-mu}
 \Delta\mu = \frac{1}{\sigma_{\rm{D}}}\int j_{1x} dx \equiv \frac{1}{\sigma_{\rm{D}}}J_{1x}\; .
\end{equation}
Notice that the integrated current $J_{1x}$ is exactly the object entering the reciprocity relation of Eq.~\eqref{eq_rec}. In the present case, by using Eq.~\eqref{eq_j_1x}, and performing the integration, we find (see Appendix~\ref{app_Jk}):
\begin{equation}
J_{1x} = - \frac{\gamma j_0^ye^{-\frac{L}{\lambda_{\rm s}}}}{\lambda_{\alpha}\sqrt{1+r_{\rm{ext}}^2}(r_{\rm{s}} + \sqrt{1+r_{\rm{ext}}^2})}\; .
\label{eq_Jx}
\end{equation}
As in the charge-to-spin conversion case, Sec.~\ref{subsec_Eperp},
even if $\tau_{\rm{ext}}^{-1}\rightarrow0$, the magnetoelectric effect does not vanish.  In Fig.~\ref{fig_SL_Jy}(b) we show $J_{1x}$ (dashed green line) as a function of  $r_{\rm{ext}}$.  In agreement with Eq.~\eqref{eq_rec}, this curve coincides with the corresponding curve in Fig.~\ref{fig_SL_Jy}(a).

\section{Local charge currents due to  the spin-galvanic effect in a finite  lateral geometry }
\label{sec_red}
In Sec.~\ref{subsec_Rcoupled} we have shown that the non-local spin injection in a system without ESR, $r_{\rm{ext}}=0$, generates a distribution of local transverse charge currents which integrate to zero [see Fig.~\ref{fig_red_mu}(a)]. Since the system was infinite in the $y$ direction, such local currents flow in the $y$ direction but do not depend on $y$. In contrast, if the system is finite in the lateral $y$ direction, then the component $j_y$ of the charge current density has to vanish at the lateral edges, and one expects a more complicated current pattern. 

Here, we compute the local distribution of the charge current density and the electrochemical potential in a system of finite width $W$. We assume  that the system has sharp boundaries at $y=\pm W/2$, and consider a particular case in which the injected spin current is polarized in the $x$ direction, $j^a_0 = \delta^{ax} j^x_0$ in Eq.~\eqref{eq_bc_j0}. 

In order to find the redistribution of the electrochemical potential, we need solve the charge continuity equation, $\partial_k j_k=0$, with $j_k$ of Eq.~\eqref{eq_jk_jk1}. This reduces to solving the Laplace equation for $\mu({\bf r})$ with the BC of zero $j_y$ at the boundaries $y=\pm W/2$.  The corresponding  boundary problem takes the following form:
\begin{equation}
\begin{split}
\label{eq_bpm}
\partial_x^2 \mu (x, y) +&\  \partial_y^2 \mu (x, y) = 0\; , \\
\sigma_{\rm{D}} \partial_y \mu (x,y) &|_{y = \pm \frac{W}{2}} = j_{1y} (x)\; ,
\end{split}
\end{equation}
where the second equation  corresponds to  zero charge current at the boundary, with $j_{1y}(x)$ from  Eq.~\eqref{eq_anom_cc1}  and plotted in Fig.~\ref{fig_red_mu}(a). Notice that the latter has two different contributions: the interfacial  $j_y^{\rm{int}}$ and the bulk one $j_y^{\rm{bulk}}$.

The boundary problem of Eq.~\eqref{eq_bpm} can be solved following the same procedure  used in Ref.~\cite{borge2019boundary}. Here, we  present  the result  in Fig.~\ref{fig_red_mu}(b). The color plot  shows the electrochemical potential, whereas the streamlines are the corresponding local charge current densities of Eq.~\eqref{eq_jk_jk1}. Interestingly, near the interface, where the term $j_y^{\rm{int}}$ of Eq.~\eqref{eq_anom_cc1} is finite,  the currents on both sides of the barrier tend to cancel it.   In the Rashba conductor, the spatial distribution is more complicated and follows the  $j_y^{\rm{bulk}}$ spatial behavior.

We explore here only the case $r_{\rm{ext}} = 0$. However, in the case of finite $r_{\rm{ext}}$ one expects a qualitatively similar  behavior of the current patterns.  The only difference is that, in that case, the integrated charge current would be finite in accordance with  Eq.~\eqref{eq_Jy}.

\section{Conclusions}
\label{sec_conclusions}
In summary, we present an exhaustive analysis of non-local magnetoelectric effects in a system with an inhomogeneous linear-in-momentum SOC. Our study is based on the SU(2)-covariant drift-diffusion equations with an additional term describing the spin relaxation due to extrinsic processes. From the spin diffusion equation we obtain BC describing diffusive systems of any dimension with interfaces between conductors with different SOC and mean-free paths. One of these BC imposes the conservation of the spin current at the interface, whereas the second BC describes the jump of the spin density when an  electric field is applied in the direction parallel to the interface. In contrast, for fields perpendicular to the interface, the second BC imposes the continuity of the spin density. 

With the help of these BC we explore the non-local SGE and its inverse in a two-dimensional hybrid structure consisting of a conductor without SOC adjacent to Rashba conductor. First, we analyze the inverse SGE, i.e., the conversion of a charge current into a spin density. When the field is applied parallel to the interface between the two conductors and in the absence of ESR, the spin induced in the Rashba conductor does not diffuse into the normal conductor. However, for a finite $r_{\rm{ext}}$, a finite SH current appears and leads to a  spin density  diffusing  into the normal conductor.   In the case in which the field is applied perpendicular to the interface, the situation is rather different. In this case, the spin generated via the local inverse SGE always diffuses into the normal conductor.  

We also study the reciprocal  effect, \textit{i.e.}, the SGE which is based on the spin-to-charge conversion. For a system with a 1D spatial inhomogeneity, we obtain  from the spin diffusion equation a direct proportionality between the local spin induced by the inverse SGE and the spatially integrated charge current induced by the direct SGE [Eq.~\ref{eq_rec}]. This relation leads to a complete reciprocity between these two observables, and we use it to study the non-local SGE in the same setup. 
Finally, we compute the  local currents and redistribution of the electrochemical potential, induced by the SGE in a system of finite lateral dimensions without ESR. 

Our results are relevant for  experiments on non-local magnetoelectric effects in  hybrid structures which combine regions with different strengths of SOC, such as semiconducting~\cite{luengo2017current}, metal-insulator~\cite{kim2017evaluation}, and van der Waals heterostructures~\cite{geim2013van}. In the latter case it is possible to build stacks of 2D materials, as for example graphene, such that the regions adjacent  to  a different  material, for instance transition metal dichalcogenides, may exhibit sizable SOC~\cite{safeer2019room, ghiasi2019charge, li2019electrical, benitez2019tunable, hoque2019all}. In such structures the SOC field is inhomogeneous in space and the electronic  transport is governed by the effects discussed in this work.

\section*{Acknowledgements}
We acknowledge funding by the Spanish Ministerio de Ciencia, Innovaci\'on y Universidades (MICINN) (Projects No.  FIS2016-79464-P and No. FIS2017-82804-P) and by Grupos Consolidados UPV/EHU del Gobierno Vasco (Grant No. IT1249-19). 

\appendix
\section{Inverse spin-galvanic effect: spatial dependence of the spin density}
\label{app_cs}
Here, we present the explicit form of the solution of the boundary problem solved in Sec.~\ref{sec_cs}. For an electric field applied parallel to the interface (Sec.~\ref{subsec_Epa}), one needs to solve  Eq.~\eqref{eq_diff_N}  in the normal conductor and Eqs.~\eqref{eq_sde_1x} and~\eqref{eq_sde_1z} in the Rashba region together with  the BC [Eq.~\eqref{eq_bc_Epa}]. The solution for the spin densities is:
\begin{widetext}
\begin{equation}
\label{eq_app_Epa}
\begin{split}
&\frac{S^x (x)}{S_{\rm{EE}}^x} = \frac{ \Theta(x)}{1+r_{\rm{ext}}^2} - \frac{r_{\rm{ext}}^2}{1+r_{\rm{ext}}^2} \frac{\Theta(x) \operatorname{\mathbb{I}m} \left\{ \left( \kappa^* + a^* (1 + r^2_{\rm{s}} + \kappa^*r_{\rm{s}}) \right) e^{-\frac{\kappa x}{\lambda_{\alpha}}} \right\} + \Theta(-x) \operatorname{\mathbb{I}m} \{ r_{\rm{s}} \kappa^* a + a |\kappa|^2 + \kappa |a|^2 \} e^{\frac{x}{\lambda_{\rm{s}}}}}{\operatorname{\mathbb{I}m} \left\{  [ a (\kappa + r_{\rm{s}}) - 1 ] (a^* + \kappa^* + r_{\rm{s}}) \right\} } \; , \\
&\frac{S^z (x)}{S_{\rm{EE}}^x} = \frac{r_{\rm{ext}}^2}{1+r_{\rm{ext}}^2} \frac{\operatorname{\mathbb{I}m} \left\{ \left( r_{\rm{s}} \kappa |a|^2 + \kappa a^* \right) \left( \Theta(x) e^{-\frac{\kappa x}{\lambda_{\alpha}} } + \Theta(-x) e^{\frac{x}{\lambda_{\rm{s}}}} \right) \right\}}{\operatorname{\mathbb{I}m} \left\{  [ a (\kappa + r_{\rm{s}}) - 1 ] (a^* + \kappa^* + r_{\rm{s}}) \right\} }\; , \\
\end{split}
\end{equation}
\end{widetext}
with $r_{\rm{s}} = \lambda_{\alpha}/\lambda_{\rm{s}}$, and:
\begin{equation}
\label{eq_a_kappa}
\begin{array}{cc}
     a = \frac{2\kappa }{\kappa^2 - \left( 2 + r_{\rm{ext}}^2 \right)}\; , &\ \kappa^2 = - \left( \frac{1}{2} - r_{\rm{ext}}^2 \right) + \frac{\im}{2} \sqrt{7 + 16r_{\rm{ext}}^2}\; .
\end{array}
\end{equation}
In the main text, for Eq.~\eqref{eq:sxz_para} we use the value of $\kappa_0$ which equals to the  $\kappa$ defined above when $r_{\rm{ext}}=0$.\section{Spin-galvanic effect: spatial dependence of the spin density}
\label{app_sc}
Here, we present the explicit form of the solution of the boundary problem solved in Sec.~\ref{sec_sc}.   Specifically, one needs to solve Eqs.~\eqref{eq_diff_N}  in the normal conductor and Eqs.~\eqref{eq_sde_1x}--\eqref{eq_sde_1y} in the Rashba region when ${\bf E}=0$.  At the boundary between the two regions, $x=0$, we impose the continuity  of the spin currents and spin densities, and at  $x=-L$ the condition of Eq.~\eqref{eq_bc_j0}. 
When the injected current $j_0^a$  is polarized in $a=x$ or $z$ directions,  Sec.~\ref{subsec_Rcoupled}, the solution reads as
\begin{widetext}
\begin{equation}
\label{eq_app_S0xz}
\begin{split}
&S^x (x) = \frac{\lambda_{\alpha}}{D} \Bigg\{ \frac{\operatorname{\mathbb{I}m} \left\{\left[ (j_0^x a - j_0^z) \left( r_{\rm{s}} + \kappa \right) - (j_0^x + j_0^z a) \right] e^{-\left( \frac{\kappa^* x}{\lambda_{\alpha}} + \frac{L}{\lambda_{\rm{s}}} \right)} \right\} }{\operatorname{\mathbb{I}m} \left\{ [ a (\kappa + r_{\rm{s}}) - 1 ] (a^* + \kappa^* + r_{\rm{s}}) \right\}} \Theta(x) \\
&+ \left[ \frac{\operatorname{\mathbb{I}m} \left\{ j_0^x a \left( r_{\rm{s}} + \kappa \right) - j_0^z \left( \kappa + a \right) \right\} }{\operatorname{\mathbb{I}m} \left\{ [ a (\kappa + r_{\rm{s}}) - 1 ] (a^* + \kappa^* + r_{\rm{s}}) \right\}} e^{\frac{x-L}{\lambda_{\rm{s}}}} + \frac{j_0^x}{2r_{\rm{s}}} \left( e^{-\frac{|x+L|}{\lambda_{\rm{s}}}} - e^{\frac{x-L}{\lambda_{\rm{s}}}} \right) \right] \Theta(-x) \Bigg\} \; , \\
&S^z (x) = \frac{\lambda_{\alpha}}{D} \Bigg\{ \frac{\operatorname{\mathbb{I}m} \left\{\left[ (j_0^x |a|^2 - j_0^z a^*) \left( r_{\rm{s}} + \kappa \right) - (j_0^x a^* + j_0^z |a|^2) \right] e^{-\left( \frac{\kappa^* x}{\lambda_{\alpha}} + \frac{L}{\lambda_{\rm{s}}} \right)} \right\} }{\operatorname{\mathbb{I}m} \left\{ [ a (\kappa + r_{\rm{s}}) - 1 ] (a^* + \kappa^* + r_{\rm{s}}) \right\}} \Theta(x) \\
&+ \left[ \frac{\operatorname{\mathbb{I}m} \left\{a\left[ j_0^z \left( r_{\rm{s}} + \kappa^* \right) - j_0^x \left( \kappa^* a^* - 1 \right) \right] \right\} }{\operatorname{\mathbb{I}m} \left\{ [ a (\kappa + r_{\rm{s}}) - 1 ] (a^* + \kappa^* + r_{\rm{s}}) \right\}} e^{\frac{x-L}{\lambda_{\rm{s}}}} + \frac{j_0^z}{2r_{\rm{s}}} \left( e^{-\frac{|x+L|}{\lambda_{\rm{s}}}} - e^{\frac{x-L}{\lambda_{\rm{s}}}} \right) \right] \Theta(-x) \Bigg\} \; , \\
\end{split}
\end{equation}
\end{widetext}
with $r_{\rm{s}} = \lambda_{\alpha}/\lambda_{\rm{s}}$, and $a, \kappa$ from Eq.~\eqref{eq_a_kappa}.
On the other hand, when the injected spin current  is polarized in the $y$ direction (Sec.~\ref{subsec_Rcoupled}), one  obtains
\begin{widetext}
\begin{equation}
\label{eq_app_S0y}
\begin{split}
S^y (x) &= \frac{j_0^y \lambda_{\alpha}}{D} \Bigg[ \frac{e^{- \left( \sqrt{1+r_{\rm{ext}}^2} \frac{x}{\lambda_{\alpha}} + \frac{L}{\lambda_{\rm{s}}} \right)}}{r_{\rm{s}} + \sqrt{1+r_{\rm{ext}}^2}} \Theta(x) + \Bigg( \frac{r_{\rm{s}} - \sqrt{1+r_{\rm{ext}}^2} }{2r_{\rm{s}}( r_{\rm{s}} + \sqrt{1+r_{\rm{ext}}^2})} e^{\frac{x-L}{\lambda_{\rm{s}}}} + \frac{1}{2r_{\rm{s}}} e^{-\frac{|x+L|}{\lambda_{\rm{s}}}} \Bigg) \Theta(-x) \Bigg] \; . \\
\end{split}
\end{equation}
\end{widetext}

\section{Integrated charge current density }
\label{app_Jk}
Here, we derive the expressions for the spatially integrated charge current density used  in Sec.~\ref{sec_sc} [Eqs.~\eqref{eq_Jy} and~\eqref{eq_Jx}], but for linear-in-momentum SOC of any kind. 
The spatial variation of the SOC is a step-like function, and therefore the SU(2) field of Eq.~\eqref{eq_Fij} has a component localized right at the interface, $x=0$, and another one homogeneous inside the Rashba conductor. 
Correspondingly, the charge current density, $j_{1k}$ in Eq.~\eqref{eq_jk_jk1},  has also an  interfacial and a bulk contribution:
\begin{equation}
\begin{split}
j_{1k} = -\gamma \Big[ \delta(x) \Big( \mathcal{A}_i^a \delta^{kx} - \mathcal{A}_k^a \delta^{ix} \Big) + \Theta(x) \mathcal{A}_k^c \mathcal{A}_i^b \varepsilon^{cba} \Big] j_i^a \; . 
\end{split}
\end{equation}  
Integrating this equation over $x$ gives:
\begin{equation}
\label{eq_Jk_0}
J_{1k} = -\gamma \Bigg[ \Big( \mathcal{A}_i^a \delta^{kx} - \mathcal{A}_k^a \delta^{ix} \Big) j_i^a |_0 + \int_{0}^{\infty} \mathcal{A}_k^c \mathcal{A}_i^b \varepsilon^{cba} j_i^a dx \Bigg]\; .
\end{equation}
On the other hand, we can also integrate the continuity equation~\eqref{eq_cont2}  over the semi-infinite Rashba conductor:
\begin{equation}
\label{eq_cc_int}
\begin{split}
\int_{0}^{\infty} \partial_x j^a_x dx + \int_{0}^{\infty} \mathcal{A}_k^c j_k^b &\varepsilon^{cba} dx = -\int_{0}^{\infty} \frac{S^a}{\tau_{\rm{ext}}} dx\; , \\
&\downarrow \\
 j^a_x |_{0}^{\infty} + \int_{0}^{\infty} \mathcal{A}_k^c j_k^b \varepsilon^{cba}& dx = -\int_{0}^{\infty} \frac{S^a}{\tau_{\rm{ext}}} dx\; ,
\end{split}
\end{equation}
where in the second line we have used the fact that $j^a_x |_{\infty} = 0$.  
Substitution of Eq.~\eqref{eq_cc_int} into  Eq.~\eqref{eq_Jk_0} gives:
\begin{equation}
\begin{split}
J_{1k} &= -\gamma \Bigg(\mathcal{A}_i^a \delta^{kx} j_i^a |_0 - \mathcal{A}_k^a \int_{0}^{\infty} \frac{S^a}{\tau_{\rm{ext}}} dx \Bigg)\; .
\end{split}
\end{equation}
This expression is a general result for the integrated  current in any hybrid structure composed of a normal and a linear-in-momentum SOC conductor with an  interface at $x=0$.

\bibliography{Poy_bib}

\end{document}